# A Survey On Dynamic Spectrum Access Techniques For Cognitive Radio

Anita Garhwal[1] and Partha Pratim Bhattacharya[2]

Department of Electronics and Communication Engineering
Faculty of Engineering and Technology
Mody Institute of Technology & Science (Deemed University)
Lakshmangarh, Dist. Sikar, Rajasthan,
Pin – 332311, India
[1]anitagarhwal@gmail.com
[2]hereispartha@gmail.com

## ABSTRACT

*Cognitive radio (CR) is a new paradigm that utilizes the available spectrum band. The key characteristic of CR system is to sense the electromagnetic environment to adapt their operation and dynamically vary its radio operating parameters. The technique of dynamically accessing the unused spectrum band is known as Dynamic Spectrum Access (DSA). The dynamic spectrum access technology helps to minimize unused spectrum bands. In this paper, main functions of Cognitive Radio (CR) i.e. spectrum sensing, spectrum management, spectrum mobility and spectrum sharing are discussed. Then DSA models are discussed along with different methods of DSA such as Command and Control, Exclusive-Use, Shared Use of Primary Licensed User and Commons method. Game-theoretic approach using Bertrand game model, Markovian Queuing Model for spectrum allocation in centralized architecture and Fuzzy logic based method are also discussed and result are shown.*

## 1. INTRODUCTION TO COGNITIVE RADIO

The radio frequency is a limited natural resource and getting enabled day by day due to growing demand of the wireless communication applications. To operate on a specific frequency band license are needed. The use of radio spectrum in each country is governed by the corresponding government agencies. In conventional technique each user is assigned a license to operate in a certain frequency band .Most of the time spectrum remains unused and it is also difficult to find it. The allocated spectrum has been not utilized properly; it varies with time, frequency and geographical locations. Thus to overcome the spectrum scarcity and unutilized frequency band, a new communication techniques cognitive radio (CR) and dynamic spectrum access (DSA) is introduced. CR network provides efficient utilization of the radio spectrum and highly reliable communication to users whenever and wherever needed. DSA technology allows unlicensed secondary system to share the spectrum with licensed primary system [1]-[2].

Joe Mitola and Gerald Maguire introduced the term CR in 1999 [3]. According to Federal Communications Commission (FCC), CR is a radio or system that senses surrounding environment and dynamically adjust its radio parameters to communicate efficiently [4]. Cognitive radio system architecture is shown in Figure 1 [5].

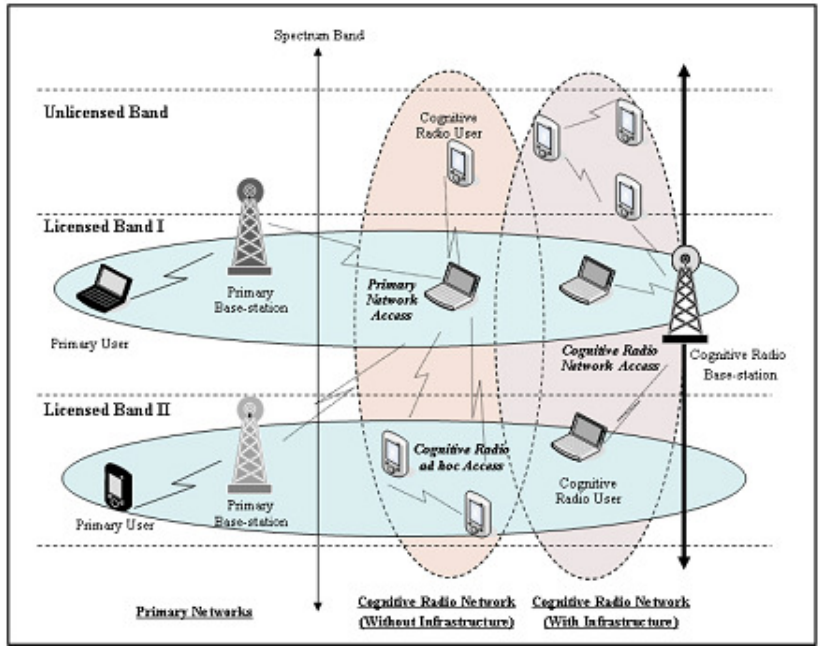

Figure 1. Cognitive radio system architecture

Efficient utilization can be improved by allowing a secondary user (SU) to utilize a licensed band when primary user (PU) is absent. So detection of spectrum hole is important as shown in Figure 2[1].

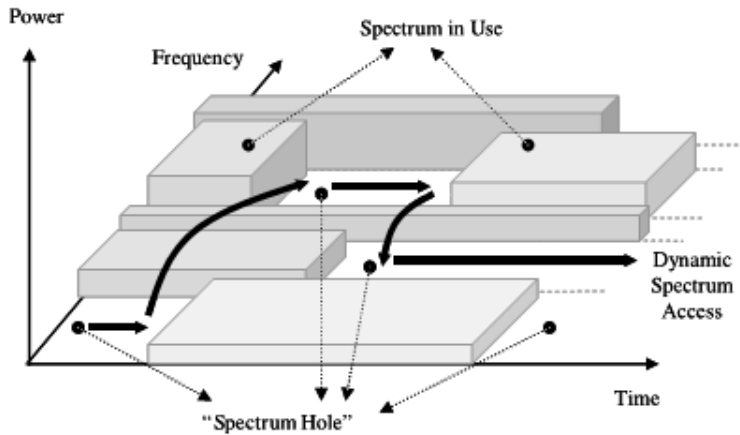

Figure 2. Spectrum hole concept

In CR technology PU are the users who have higher priority to use a specific spectrum. On the other hand, SU are the users who have lower priority, it uses the spectrum without causing harmful interference to PU.

Cognitive Radio has two major characteristics:

• Cognitive capability

Cognitive capability enables the cognitive radio to sense the information from the radio environment in order to find out the unused radio spectrum at a specific time or location. Then the appropriate portion will be selected for the communication without causing harmful interference to the other users [6].

Cognitive cycle requires adaptive operation in open spectrum access. Three major parts of the cognitive cycle are: spectrum sensing, spectrum analysis, spectrum decision as shown in Figure 3 [3].

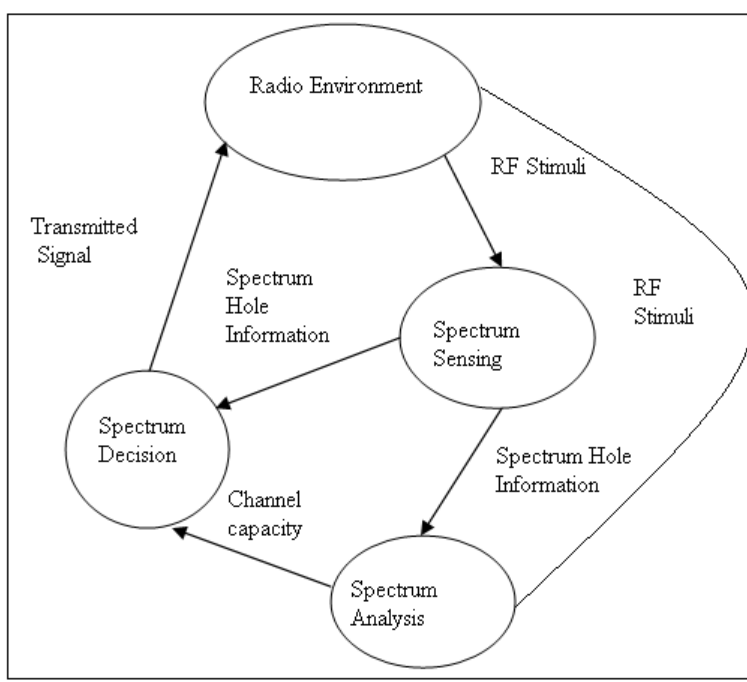

Figure 3. Cognition cycle

- Spectrum Sensing: Determine which portion of spectrum is available and detect the presence of licensed users and spectrum hole.

- Spectrum Analysis: Performs estimation of spectrum hole through spectrum sensing.

- Spectrum Decicion: A CR determines the channel capacity, spectrum hole information along with data rate and bandwidth of the transmission. Appropriate spectrum band is chosen for transmission of the signal.

- Reconfingurability

Reconfigurability implies the radio spectrum to be dynamically changes the functions according to surrounding i.e. cognitive radio can change the radio frequency, transmission power, modulation scheme, communication protocol without any modification of the hardware environment.

## 1.1. Main Functions of CR

Cognitive radio includes four main funcational blocks [6]:

### 1.1.1 Spectrum Sensing

It determies the available spectrum band and detect the primary licensed users. So spectrum sensing is significant in CRs in avoiding collision with the licensed user and improving the licensed spectrum utilization efficiency.To enhance the detection probability many signal detection technique can be used in spectrum sensing, these are classified as:

i) Primary Transmitter Detection (Non cooperative detection): It refers to determination of a signal from a primary transmitter whether it it present or not. This technique utilizes the detection of the signal from a primary transmitter. Basic hypothesis model for transmitter detection can be defined as follows [7]:

$$x(t) = \begin{cases} n(t) & H_0, \\ hs(t) + n(t) & H_1, \end{cases} \quad (1)$$

where x(t) is received signal, s(t) is the primary users's transmitted signal, n(t)is the AWGN(Additive White Gaussian Noise) and channel's amplitude gain is h. $H_0$ is a null hypothesis, $H_1$ is an alternative hypothesis.

According to hypothesis model three schemes are used for transmitter detection in cognitive radio[8].

• Matched Filter Detection

A matched filter is a optimal detector in stationary Gaussian noise which maximizes the received signal-to-noise ratio (SNR). The block diagram is shown in Figure 4 [1].

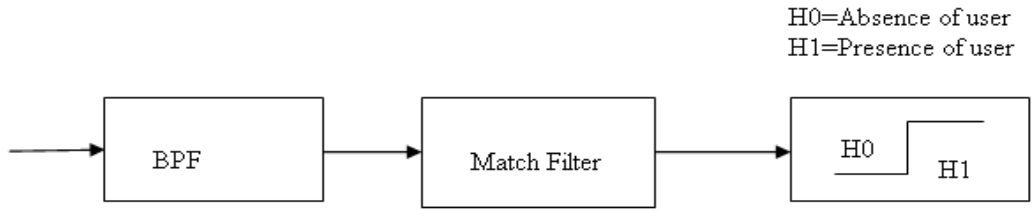

Figure 4. Block diagram of matched filter

The operation of matched filter detection is expressed as[1]:

$$Y[n] = \sum_{k=-\infty}^{\infty} h[n-k]x[k] \quad (2)$$

Where 'h' is convolved with 'x' for maximizing the SNR, the impulse response of matched filter and reference signal are matched. When information from the primary users is known then matched filter detection is useful.

This technique has the advantege that it requires less detection time because it requires less time for higer processing gain. It has disadvantage that priror knowledge of primary signal such as pulse shape, modulation type is needed.

• Energy Detection

In this method detection of the primary signal is based on the sensed energy[1]. If prior knowledge of the PU signal is unknown, the energy detection method is optimal for detecting any signals. In this approach, the radio-frequency (RF) energy in the channel or the received signal strength indicator is measured to determine whether the channel is idle or not.

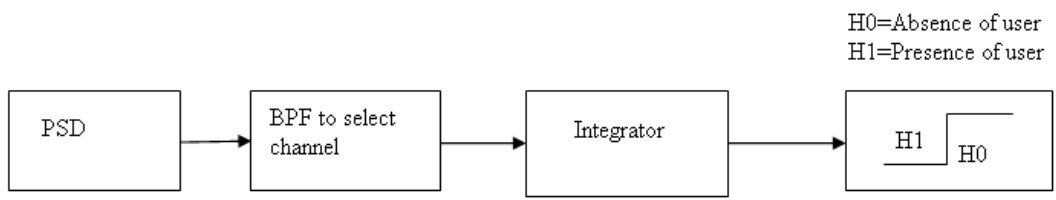

Figure 5. Block diagram of energy detector

Figure 5 shows block diagram of energy detector [1]. In this method, signal is passed through band pass filter to select channel and is integrated over time interval. Finally the output of the integrator is compared with a threshold to determine whether primary user is present or not. Based on he channel conditions the threshold value can set to be fixed or variable. Analytically, signal detection can be reduced to a simple identification problem, formalized as a hypothesis test [1] .

Disadvantages of the scheme are as follow - (1) The threshold used in energy selection depends on the noise variance. (2) Inability to differentiate the interference from other secondary users sharing the same channel and the PU. (3) It has poor performance under low SNR conditions.This is because the noise variance is not accurately known at the low SNR, and the noise uncertainty may render the energy detection useless.

• Cyclostationary Feature Detection

To identify the received primary signal in the presence of primary users it exploits periodicity of modulated signals couple with sine wave carriers, hopping sequences, cyclic prefixes etc. Block diagram of cyclostationary feature is shown in Figure 6 [9].

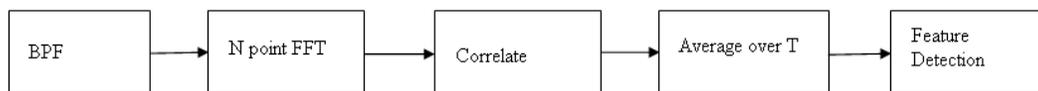

Figure 6. Cyclostationary feature detector block diagram

This technique is robust in discriminating in noise so it performs better than energy detector.It have demerit that it need more computational complexity and longer observation time.

ii) Cooperative Detection

In this technique for detection of primary user multiple CR users are incorporated. In primary transmitter detection technique there was a hidden terminal problem exist while having a good line-of-sight to receuver CR transmitter may not able to detect the detect the transmitter due to shadowing as shown in Figure7 [6].

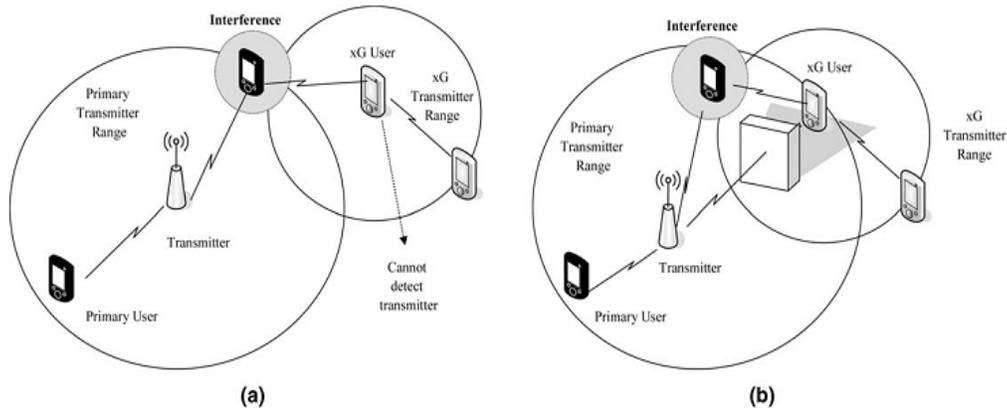

Figure 7. Transmitter detection problem: (a) Receiver uncertainty and (b) Shadowing uncertainty

Cooperative sensing techniques are classified as (1) Centralised Coordinated (2) Decentralised Coordinated (3) Decentralised Uncoordinated.

iii) Interference -Based Detection

Interference is actually happen at receivers, it is regulated in trans-centric way and is controlled at the transmitter through the radiated power, location of individual transmitters. Interference temperature model is shown in Figure 8 [10].

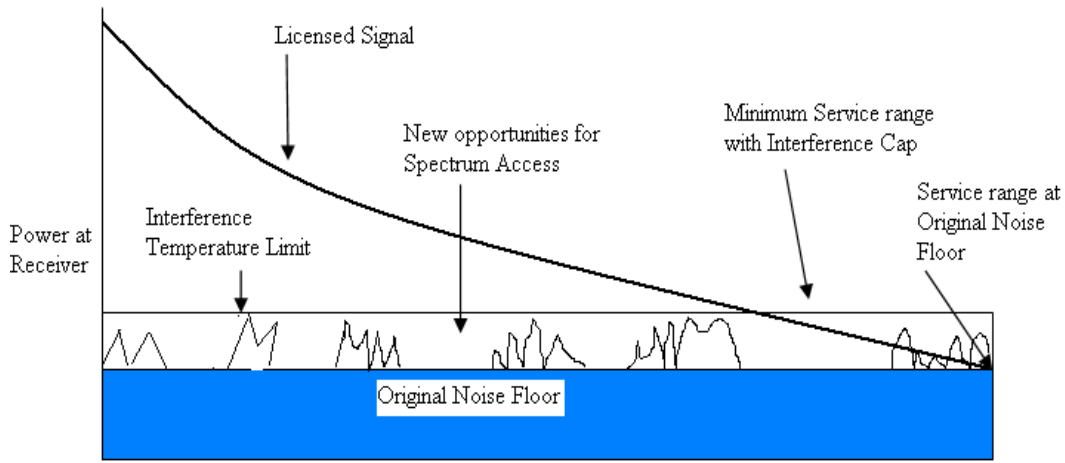

Figure 8. Interference temperature model

### 1.1.2. Spectrum Management

In this technique based on the availability of the spectrum and other policies, CR user allocates the best available spectrum band to achieve high quality of service requirement. There are two technique for spectrum managemt:

• Spectrum analysis: In this technique each spectrum hole should be characterized considering not only the time-varying radio invironment and but also the primary user activity.

• Spectrum Decision: when all the analysis of spectrum band is done, appropriate spectrum band is be selected for the current transmission considering the QoS requirements and the spectrum characteristics. According to user requirement the data rate, bandwidth is determined then according to decision rule appropriate spectrum band is chosen.

### 1.1.3. Spectrum Mobility

In this process CR changes its frequency of operations to use the spectrum in dynamic manner to operate in the best available frequency band.When primary user appears, current channel condition become worse so spectrum mobility arises. Due to this spectrum mobility spectrum handoff arises.

### 1.1.4. Spectrum Sharing

It is the major challenge of the open spectrum usage. There is a coordinayed access with other users.This technique consist of five major steps: spectrum sensing, spectrum allocation, spectrum access, transmitter-receiver handshake, spectrum mobility.

## 2. INTRODUCTION TO DYNAMIC SPECTRUM ACCESS

Dynamic Spectrum Access (DSA) is necessary in order to allocate the available bandwidth in an efficient and effective manner .The legacy methods of assigning different fixed bandwidths to different system are not producing the full benefits of having dynamically shared bandwidths for different systems only as and when they need them. Dynamic spectrum access can help to minimize unused spectral bands or white spaces [11]. Cognitive radio have one important property that it must not interfere with licensed band. When the primary user wants to start transmission, the CR enabled device free that band and switch to another free band. This technique of the dynamically accessing the unused bands for proper utilization is known as DSA [12].

## 3. DYNAMIC SPECTRUM ACCESS MODELS

Current spectrum management policy is based on the fixed allocation of the radio resources, little sharing of radio spectrum which results in spectrum shortages. In comparison to the static spectrum access, dynamic spectrum access is widely used and having various approaches and applications.

## 3.1. Different Approaches of Dynamic Spectrum Access Models

Dynamic spectrum access models are four types as shown in Figure 9 [13].

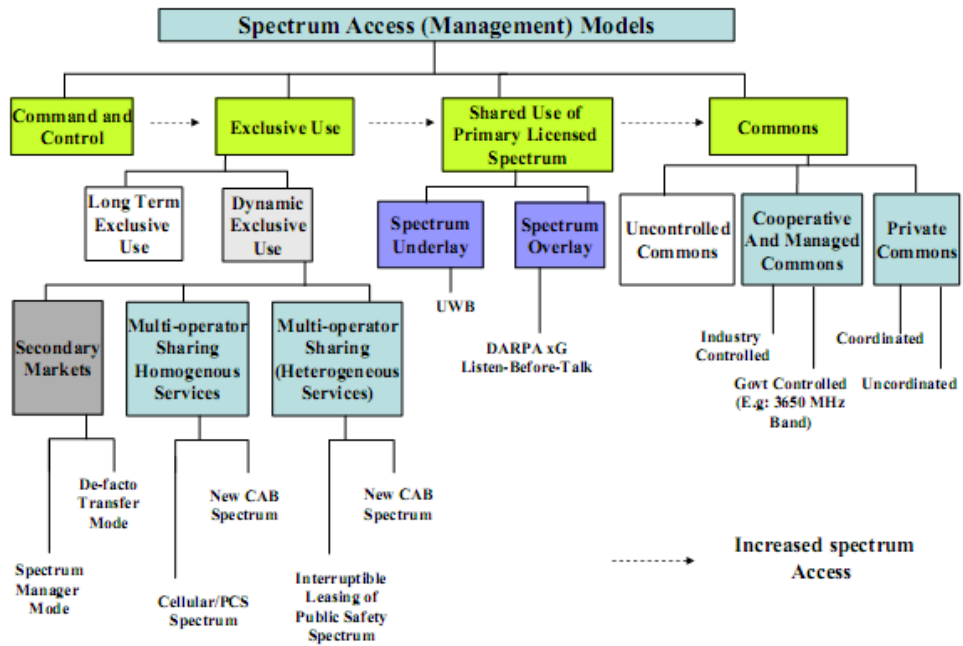

Figure 9. Taxonomy of Spectrum access models

### 3.1.1. Command and Control

In this model, the regularity body clearly governs the rule for spectrum uses and assigns an entity for use. As there are no competing interest or entities allowed seeking access to spectrum, ownership assignment does not use any market mechanism. The devices have exclusive and nearly eternal access to the spectrum band when they assigned the spectrum.

### 3.1.2. Exclusive-Use

This spectrum maintains the basic structure of the current spectrum regulation policy where spectrum bands are licensed to serve for exclusive use. The main objective is introducing flexibility in spectrum allocation and usages. Two approaches have been proposed in this model: (a) Long-term Exclusive-Use (b) Dynamic Exclusive-Use.

(a) Long-term Exclusive- Use

A model that manages spectrum using space, frequency and type-of-service dimensions and guarantees exclusive ownership with those constraints for prolonged periods of time. It has two variations. (1) Fixed Use type where license cannot change the type of service, technology during the license period. (2) Flexible Use type where the license can change the type of service, technology used in band during license life time.

(b) Dynamic Exclusive-Use

This model manages spectrum in finer scale of time, space, frequency and use dimensions so at any given point in space and time, only one operator has exclusive right to the spectrum but the identity of the owner and type of use can change. It has three variants:

(i) Secondary Markets

It is a general mechanism in which license alteration is possible. Non real time Secondary markets: To promote the secondary markets for transfer of spectrum usage rights, regulatory authorities in USA and UK namely FCC and Ofcom have recently explored measures.

FCC offered up two modes of spectrum rights transfer, also called spectrum leasing.

• Spectrum Manager

In this mode, the primary licensee act as a spectrum manager and the lessee with transferred license report back to the licensee for matters relating to the license. The final legal ownership, the responsibility of ensuring lessee's compliance with the original license terms and duties of periodic reporting to the FCC still remain with the licensee.

• De Facto Transfer

Here, the license rights are effectively transferred in total to the lessee. The responsibility of reporting back to FCC and conforming to service and interference rules of the license is entirely lessees [13]. FCC lease approval is a non real time, slow process. [14]-[15].

(ii) Homogeneous Multi-operator Sharing via Real-time Secondary Market
We will describe this model with respect to current commercial cellular services in USA. The frequency bands used are 800 MHz and broadband PCS 1900 MHz as shown in Figure 10[13].

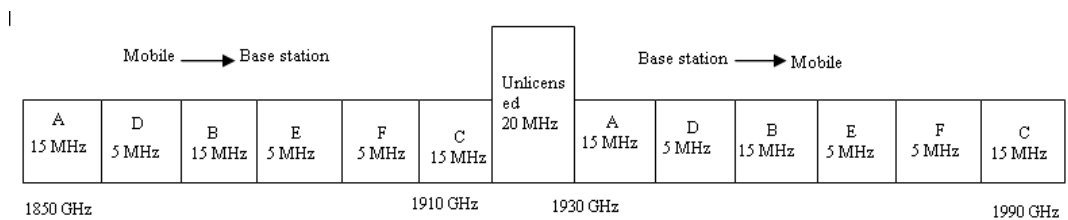

Figure 10. PCS band plan in USA in 1900 MHZ

The broadband PCS spectrum consists of 120 MHz of spectrum consisting of two bands each of 60 MHz: the mobile to base station 1850-1910 GHz band and base station to mobile 1930-1990 GHz bands. Each of the bands is divided into six bands (A, B, C: 15 MHz, D, E, and F: 5 MHz).
In a given geographic region, a cellular provider owns a license to a fixed block A, B of cellular and/or (A to F) PCS bands. Licensing process guarantees that the block assignment is mutually exclusive, thus guaranteeing strict spectrum silos to each provider. Each provider then deploys the base station via a network dimensioning/provisioning process and each base station site is registered with FCC in the Universal Licensing System (ULS) database [16].

23

Consider a scenario where new spectrum available for dynamic access and sharing. In this, license owners continue to have exclusive rights to their spectrum and may optimally share the spectrum. The regulatory authority will allocate a new band of spectrum that is designated a priori for dynamic sharing. We envision two entities: one a spectrum access provider (SAP) that has exclusive long term license to the spectrum but is mandated not to own any facilities and a second set of entities–the facilities or Radio Infrastructure Provider (RIP) who do not any long term licensed spectrum. The spectrum access and reversion is thus explicitly controlled by SAP and therefore, the spectrum band can be called Coordinated Access Band (CAB) [17].

The advantage of this model is that the base stations and client devices required can be designed using current technologies.

(iii) Heterogeneous Multi-Operator Sharing via Real-time Secondary Market

In this model, the spectrum is shared among different spectrum service providers such as cellular, public safety, broadcast TV on various Spatio- Temporal scale. The assumption here is that the facilities-based infrastructure (e.g.: cell towers for public safety, TV towers) is deployed in each allocated and licensed band and operated independently by different operators. For example, in the context of certain services such as public safety, average utilization can be low but in the event of emergency, service availability must be instantaneous and reliable. As such, in this case, shared usage of such spectrum must be interruptible with strict guarantees. Markus et al. [18-19] and FCC NPRM on Software Defined Radios (SDR) [20] call this the concept of interruptible or callable spectrum.

### 3.1.3. Shared-Use of Exclusive Licensed Spectrum

Same as Exclusive-Use model, this model is first to describe simultaneous shared use of spectrum wherein there is a primary licensed owner of the spectrum band and multiple secondary users opportunistically share the band. Spectrum sharing between primary and secondary users utilizes spectrum underlay and spectrum overlay approaches.

• Spectrum Underlay

The underlay approach allows primary and secondary user transmission simultaneously in the manner of ultra wideband (UWB) systems. To protect the primary users, system provides a spectral mask to secondary signals so that the interference generated by the secondary devices is below the acceptable noise floor for the primary users of the spectrum. This technique allows communication over short range [21].
The spectrum underlay breaks the Secondary-Usage and Flex-Use barriers but by its very conservative nature is not suitable for most aggressive spectrum use as it limits the type of network configurations.

• Spectrum Overlay

With the help of this technique primary and secondary user are allows transmitting simultaneously. The alternate for overlay system is that for secondary communication secondary users use their power and the remaining power to relay primary transmission [21].

Spectrum overlay also was first envisioned by Mitola [22] under the term "spectrum pooling" and later investigate by DARPA xG PPROGRAM [23] ONDER THE TERM (OSA) "opportunistic spectrum access" Spectrum overlay approach targets at spatial and temporal unused radio spectrum called white space by allowing secondary users to identify and exploit local and instantaneous spectrum availability in nonintrusive manner. Therefore exclusive knowledge about other signals in the spectrum is necessary.

### 3.1.4. Commons

One of the meanings of the word "Commons" refers to "a piece of land for common public use (such as cattle grazing)" indicating a operating model wherein nobody can claim exclusive use of a shared resource. The argument that radio spectrum is a public resource that should be equitably and fairly accessible to everyone without undue government regulation is frequently termed as "Commons" model of spectrum access or management. The commons model of spectrum access has three variants described in the following:

• Uncontrolled Commons

When a spectrum band is managed using the uncontrolled commons model the simplest and purest commons of all, no entity has exclusive license to the spectrum band. As such, anyone can have any number of devices operating in such a band and the model is therefore, often referred as "open spectrum access". The current ISM (2.4 GHz), U-NII (5 GHz) unlicensed bands represent examples such commons. Here, FCC only mandates that the devices conform to a peak transmit power [24].

• Managed Commons

It avoids the tragedy of commons by imposing a limited form of order or structure to spectrum access [25]. A commons is a resource that is owned or controlled jointly by a group of individuals or entities and it is characterized by restrictions on who uses the resource, and when and how [25] it is used. The entity or a group of entities that establishes and enforces these restrictions is the controller of the commons. As such the notion of managed commons may be consider as the true commons.

This model provides shared among group of devices, unlicensed, free use of spectrum by devices using multiple technologies. Two things are central to use of managed commons as viable spectrum access technique: (1) A good commons management protocol that encapsulates technology agnostic rules. (2) Reliable, scalable mechanisms that quantify rule on performance of participating devices (entities).

• Private Commons

Private Commons is a new concept introduced by FCC in its Second Report and Order and second further NPRM on elimination of barriers to development of Secondary markets for spectrum [15]. This concept is aimed at allowing use of advanced technologies (e.g: Frequency agile radios capable of listening to coordination channels or of spectrum sensing) that enable multiple parties to access the spectrum to be gradually employed in existing licensed bands at the discretion of license holder. It is a policy mechanism for creating a managed commons where the ultimate ownership of the licensed spectrum is still centralized with the license holder.

The private commons implemented in licensed band with licensee control helps in quality-of-service conscious users affected by overcrowding in open access, unlicensed bands. Design of standardized protocols and devices for this model has strong market opportunities.

There are two ways in which the private commons may be realized in the existing licensing paradigm [15].

• License Holder Offers Private Commons Service

In this model the licensee holder (primary licensee) maintain a customer relationship with device owner and charge a transition fee for the same. This type of model may be beneficial for end users who may find guarantees provided by the licensed band attractive in addition to unlicensed bands that may be crowded.

• License Holder Offers Spectrum Access

In this model the license holder does not maintain a customer relationship with end- users interested in access to licensed spectrum and may not own any network. It makes agreement with one or more device manufacturers, who work with license or independently to specify and develop devices to use the license holder spectrum. There is no primary user of the license.

This mode is different from the managed commons in the sense that license ownership and the terms and conditions of use are dictated by fixed rather than dynamically changing collection of entities. Currently, FCC proposes to restrict private commons to only peer-to-peer communications between devices in a flat network that does not use primary licensee's carefully engineered network [15]. As such this, mode is suitable for adhoc and mesh networks.

Private commons is a middle ground; it allows an exclusive use license holder to "get adventurous" to aggressively use its spectrum, potentially for increased revenue stream via per user periodic transactions – a form of a spot secondary market. It eliminates or minimizes the possibilities of over exploitation and provides better guarantees to end- user than in open access.

## 3.2. Game-Theoretic Approach for Spectrum Sharing in Cognitive Radio Networks

Cognitive transceiver has an ability to sense and reconfigure the own device parameter according to the outside world. With this agile property of the transceiver, spectrum utilization can be improved using by sharing the frequency spectrum band among the licensed (i.e. primary) and unlicensed (i.e. secondary) users. On considering selfish behavior of current network devices, a game-theoretic spectrum sharing criteria would be required to maximize both primary and secondary user's satisfaction [26].

When the allocated frequency spectrum is not fully utilized by primary users, then they can gain more profit by selling it to secondary users who want to utilize it opportunistically. In sharing of unutilized frequency spectrum, proper criteria would be needed. When multiple buyers and sellers exist then an oligopoly pricing model is used for sharing method.

We model the spectrum sharing as an oligopoly market where the primary user has insufficient quantity of spectrum to sell because of primary users channel usage pattern. For the secondary users, this insufficient quantity of spectrum is less attractive as secondary users want to use the frequency band keep unutilized as long as possible. Thus the preference for long sojourn-time must be considered.

• Sojourn Time Formulation

The mean sojourn time is defined as the amount of time an object is expected to spend in a system before leaving the system for good.

Alternating renewal theory suggests that remaining time of certain random variable follow [26]:

$$f_x = \int \frac{1-F(x)}{E[X]} dx \qquad (3)$$

Where x denotes the certain random variable F(x) is cumulative distribution function and E[X] is an expected value of random variable. With this theorem we can formulate remaining time of available opportunity between two consecutive sensing for secondary users as follows:

$$L_i = \int_t^{t+T_p} x \cdot \frac{1-F_{T_{OFF}^i}(x)}{E[T_{OFF}^i]} dx \qquad (4)$$

Where t denotes first sensing time among the consecutive sensing. This sojourn-time constraint is considered as a limit of the demand for channel i.

• Static Bertrand Game Model

In this model there is interaction among firms (sellers) and their customers (buyers). Buyers choose quantities at a price that is fixed by sellers. We can formulate a Bertrand game as follows. The strategy of the each channel player (PU) is to set the unit price of the quantity of the spectrum. The payoff is profit of selling spectrum to secondary users. The quantity of the spectrum (i.e., demand from the secondary users) depends upon price competition (i.e., channel quality, substitutability) and cannot exceed capacity (i.e., sojourn-time constraint). The solution of the game is set pure strategies to reach Nash equilibrium. The Profit function for each channel of primary user is given as follows:

$$p_i \times q_i = -b_i p_i^2 + (a_i + \sum_{j \neq i} c_{ij} p_j) p_i \qquad (5)$$

In game theory if each player has chosen a strategy and no player can benefit by changing his or her strategy while the other players keep their unchanged, then the current set of strategy choices and the corresponding payoffs constitute Nash equilibrium. Mathematically to derive Nash equilibrium of this game we have to solve the set of marginal profit function (i.e., partial differentiation of profit function respect to pi): ∂πi (p)/∂pi = 0 for all players of the game as follows:

$$p_i = \frac{a_i + \sum_{j \neq i} c_{ij} p_j}{2b_i} \qquad (6)$$

Thus we can obtain the Nash equilibrium under sojourn-time (demand exceeds the capacity) constraint by the demand (i.e., $q_i$) and the capacity (i.e., $L_i$) being equal as follows:

$$p_i = \frac{a_i - L_i + \sum_{j \neq i} c_{ij} p_j}{b_i} \qquad (7)$$

### 3.3. Markovian Queuing Model for Dynamic Spectrum Allocation in Centralized Architecture

In centralized architecture a central controller is responsible for allocation of bandwidth to intended users. In order to overcome the hidden terminal problem and to obtain complete information of unused frequency (spectrum hole) sensing is considered to be decentralized. In each SU one transceiver is dedicated for control and second a SDR based which scans the availability of spectra in its vicinity and forwards the information of these spectrum holes to the master/controller in case the SUs form an infrastructure less network or to the BS in case of infrastructure based network. The bandwidth is allocated to SU based on the number of availability of unoccupied channels. The equivalent model of network of queues is as shown in Figure 11 [27].

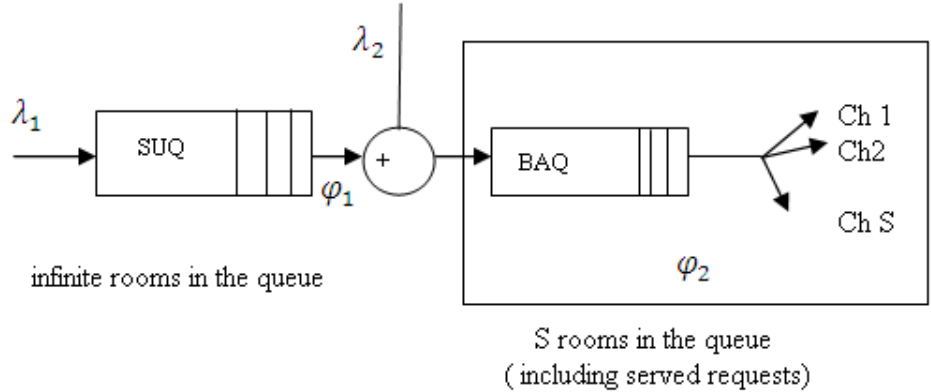

Figure 11. Queuing model for DSA in CR network [27]

These queues are special cases of stochastic processes, characterized by an arrival process of service requests, a waiting list of requests to be processed. The queue stacking all entries of SUs is referred as a secondary users queue (SUQ) and all the requests entering this queue are served on first come first serve (FCFS) basis. At any time when bandwidth needs be allocated to the SU, the Head considers both the requests from SU and the PU who need its licensed channel. Thus while distributing number of frequencies to PU and SU; the arrival rates of both the users are summed to access the frequencies with the Head. The queue so formed is referred to as bandwidth allocation queue (BAQ).

We use Markov process to analyze the queuing models. The blocking probability $P_B$ for the bandwidth request made by CR that finds all the channels with Head as occupied is given by Erlang-B formula as given below:

$$P_B \equiv P_S = \frac{\rho_2^S}{S! \sum_{i=0}^{S} \frac{\rho_2^i}{i!}} \qquad (8)$$

Where $\rho_2$ is traffic intensity for BAQ.

The blocking probability from equation (8) is plotted in Figure 12. Variation in $P_B$ is shown with respect to change in number of available channels in the system as 2, 5, 7, 10, 13, and 15. Blocking probability increases with increase in SU traffic in the network.

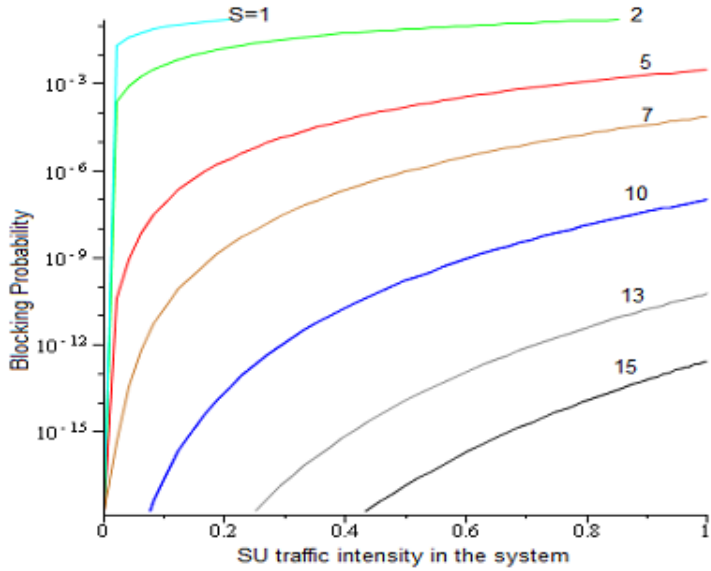

Figure12. Blocking probability ($P_B$) against SU utilization in the system (ρ1). The variation in $P_B$ is depicted with different numbers of channels (S), available with the system.

### 3.4. A Fuzzy Logic Based Spectrum Access Method

The model of the fuzzy based system is shown in Figure 13 [28].

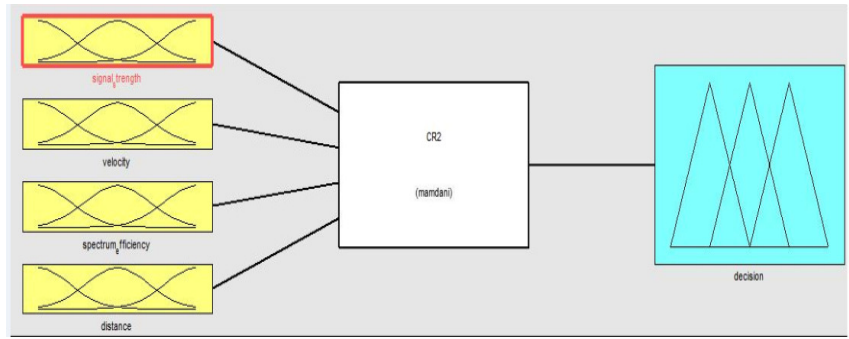

Figure 13. Fuzzy logic system

The determining input parameters are signal strength, node velocity, secondary user velocity, spectrum efficiency and distance between the primary licensed and secondary unlicensed user [28]

Some of the simulation results are shown in Figure 14 and Figure 15. It may be seen from the results that the chance of taking decision increases if the signal strength of the channel offered by primary user is high and the distance between primary and secondary users is low (Figure 14). As the velocity increases the chance of spectrum accessing is more if the distance is reasonably small (Figure 15).

International Journal of Next-Generation Networks (IJNGN) Vol.3, No.4, December 2011

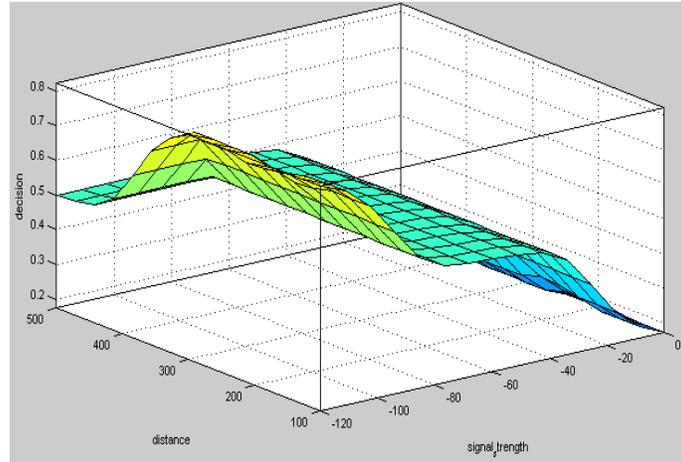

Figure 14:  Opportunistic spectrum access decision possibility (Velocity = 50 Km / hr and ratio of required spectrum to available spectrum = 0.5)

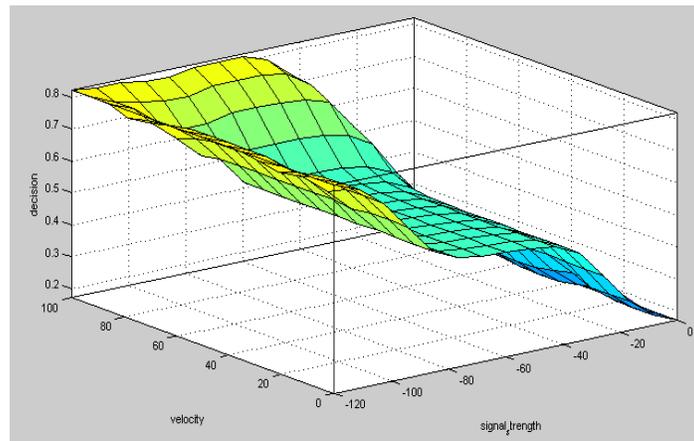

Figure 15: Opportunistic spectrum access decision possibility (distance between primary and secondary users = 50 meters and ratio of required spectrum to available spectrum = 0.5)

## **4. CONCLUSIONS**

In this paper, functions of CR are discussed for spectrum sensing with non-cooperative detection which includes matched filter, energy detection and cyclostationary fetaure detection techniques. Spectrum management includes spectrum analysis and spectrum decision property. Different specrtum access method such as command and control, exclusive-use which includes long term exclusive-use, dynamic excluisve-use are discussed. Shared use of exclusive licensed spectrum model uses spectrum underlay and overlay approaches. Commons model includes  uncontrolled, managed and private commons.  In Markovian queuing model a centralized architecture is proposed for alloction of bandwidth and to find blocking probability. Game–theoritic spectrum sharing is proposed to obtain Nash equlibrium under sojurion- time constraints using static bertrand game model. A fuzzy logic based spectrum sensing technique is also discussed in which possibility of decision increases when signal strength is weak, velocity of the node is high, spectrum efficiency is low and distance from secondary user is low.

## Authors

Anita Garhwal was born in India on July 18, 1988. She passed B. Tech in Electronics and Commu nication Engineering from Mody Institute of Technology and Science (Deemed University), Rajasthan, India. She is now a final year student of M.Tech (Signal Processing) in Mody Institute of Technology and Science (Deemed University), Rajasthan, India.

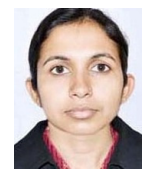

Partha Pratim Bhattacharya was born in India on January 3, 1971. He received M. Sc in Electronic Science from Calcutta University, India in 1994, M. Tech from Burdwan University, India in 1997 and Ph.D (Engg) from Jadavpur University, India in 2007. He has 15 years of experience in teaching and research. He served many reputed educational Institutes in India in various positions. At present he is working as a Professor in Department of Electronics and Communication Engineering in the Faculty of Engineering and Technology, Mody Institute of Technology and Science (Deemed University),

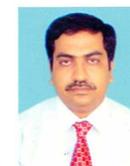

Rajasthan, India. He worked on Microwave devices and systems and mobile cellular communication systems. He has published a good number of papers in refereed journals and conferences. His present research interest includes mobile cellular communication and cognitive radio. Dr. Bhattacharya is a member of The Institution of Electronics and Telecommunication Engineers, India and The Institution of Engineers, India. He received Young Scientist Award from International Union of Radio Science in 2005. He is working as the chief editor, editorial board member and reviewer in many reputed journals.